# Proposal and Evaluation of a Practical CBCT Dose Optimization Method

**Authors:** S. Gros[1], J. Bian[2], J. Jackson[1], M. Delafuente[1], H. Kang[1], W. Small Jr[1], M. Mahesh[3]

*[1] Department of Radiation Oncology, Stritch School of Medicine, Loyola University Chicago, Maywood, IL, USA*
*[2] Department of Radiology and Medical Imaging, Stritch School of Medicine, Loyola University Medical Center, Maywood, IL, USA*
*[3] Johns Hopkins University School of Medicine, Baltimore, MD, USA*

## Abstract

Objective: To propose a CBCT dose optimization method developed with measurement equipment readily available in radiation oncology departments.

Approach: A 0.6cc air Kerma ($K_{air}$) calibrated Farmer chamber was placed at isocenter to measure the $K_{air}$ of 5 default CBCT protocols (HEAD, PELVIS, LARGE PELVIS, THORAX, SPOTLIGHT) clinically used on a VARIAN TrueBeam linac. Imaging parameters of each protocol were modified to lower the cumulative exposure either via (1) reducing the number of projections (by 25%, 50% and 75%) used by the built-in FDK reconstruction algorithm, or (2) by reducing single frame exposure by 25%, 50% and 70-75%. Cone beam Dose Index (CBDI) and weighted CBDI ($CBDI_w$) were measured in cylindrical CTDI phantoms to establish their relationship to $K_{air}$ at isocenter. Quantitative CBCT imaging quality metrics (HU linearity, HU uniformity, low contrast visibility, high contrast resolution, and spatial integrity) were evaluated for each of the modified protocols against the default protocols in a Catphan604 (PhantomLab, Greenwich NY) with a free open-source automated imaging QA analysis software. Qualitative Image quality was assessed for clinical use with Head, Thorax and Pelvic anthropomorphic phantoms.

Main results: Linear relationships between $K_{air}$ and CTDI and CTDIw indices were established for all protocols. Spatial integrity and high contrast resolution were uncorrelated with exposure ($R<0.5$) while strong correlations ($R>0.75$) were observed for low contrast visibility. HU uniformity and linearity remained stable (<40HU from baseline) for all modified protocols except for the LARGE PELVIS below the 50% single frame exposure level. Qualitative inspection of images revealed that lowering the imaging dose by 80%-50% retained clinically acceptable imaging quality with minimal interference from aliasing artifacts or noise.

Significance: The default CBCT protocols available on Truebeam linacs were optimized for imaging dose reduction. The proposed imaging dose optimization method can be easily implemented by radiotherapy physicists without access to, or expertise with diagnostic physics equipment.

## Introduction

Image-guided radiotherapy (IGRT) has become the standard of care in external beam radiotherapy (RT) [1], enabling precise tumor targeting through daily imaging. Cone-beam computed tomography (CBCT), a cornerstone of IGRT, provides volumetric imaging that facilitates accurate patient positioning and adaptive radiotherapy. However, the cumulative imaging dose from CBCT can rival the therapeutic dose by the end of treatment, particularly for highly fractionated regimens [2]. This is especially concerning for sensitive populations such as pediatric patients [3] and long-term cancer survivors [4-7], where the risks of secondary malignancies and other radiation-induced effects are significant.

Managing imaging doses during IGRT presents unique challenges. The methods proposed by AAPM and IAEA for dose output measurement [8] and optimization are often impractical for routine use due to their complexity and time requirements. Additionally, radiotherapy physicists may lack access to diagnostic-quality equipment or the expertise required for imaging dose optimization. Further complicating the issue, vendors of IGRT equipment rarely provide tools to facilitate systematic imaging dose management, leaving many clinics without practical solutions for routine dose optimization , relying on default manufacturers provided CBCT protocols [9].

To address these gaps, we propose a practical CBCT dose optimization method for Varian TrueBeam (Varian Medical Systems, Palo Alto, CA) medical linear accelerators (Linacs), developed using measurement equipment readily available in most radiation oncology departments. This method involves modifying imaging parameters to reduce exposure while maintaining clinically acceptable image quality, enabling radiotherapy physicists to implement dose optimization without relying on diagnostic physics equipment or extensive training.

The objective of this study is to evaluate the feasibility of the proposed method by establishing relationships between dose indices, imaging quality, and clinical applicability for optimized protocols. The results provide a practical framework for reducing imaging doses during IGRT, ensuring compliance with radiological protection principles while retaining the diagnostic efficacy needed for accurate treatment delivery.

## Method

## CBCT protocols

A VARIAN TrueBeam linear accelerator (Serial Number 3394) with software version MR 2.7 XI was used for this project. The clinical protocols evaluated were the default CBCT protocols provided by Varian, including HEAD, THORAX, PELVIS, LARGE PELVIS, and SPOTLIGHT. Table 1 summarizes the default imaging parameters for these protocols, including field-of-view (FOV), tube potential (kVp), and total exposure (mAs). Imaging parameters were systematically modified in service mode to reduce total exposure through three approaches:

1. Reduction in the number of projections: Frames were reduced by 25%, 50%, and 80%, using the built-in Feldkamp-Davis-Kress (FDK) reconstruction algorithm.
2. Reduction in single-frame exposure: The mAs per frame was decreased by 25–33%, 50%, and 70–75%.
3. Combination of projections and single-frame exposure reduction: Combined adjustments were evaluated to maximize dose reduction without compromising imaging quality.

A total of 43 modified protocols were evaluated. The minimum attainable exposure levels in service mode were 18 mAs for full rotation protocols (PELVIS, LARGE PELVIS, and THORAX) and 10 mAs for half rotation protocols (HEAD and SPOTLIGHT).

*Table 1: VARIAN default protocols parameters and associated dosimetry.*

| Protocol name | Filter | Scan length (cm) | FOV (cm) | kV | No. of frames | mA | ms | Total Exposure (mAs) | Angular range (deg) | CBDI (mGy) | $CBDI_w$ (mGy) | K_air (mGy) |
|---|---|---|---|---|---|---|---|---|---|---|---|---|
| **head** | Full Fan | 21.4 | 28 | 100 | 500 | 15 | 20 | 150 | 200 | 3.3 | 3.2 | 5.3 |
| **thorax** | Half Fan | 21.4 | 49.4 | 125 | 900 | 15 | 20 | 270 | 360 | 3.7 | 5.0 | 17.8 |
| **pelvis** | Half Fan | 21.4 | 49.4 | 125 | 900 | 60 | 20 | 1080 | 360 | 13.5 | 18.1 | 64.3 |
| **Pelvis large** | Half Fan | 21.4 | 49.4 | 140 | 900 | 75 | 25 | 1688 | 360 | 29.1 | 38.3 | 132.0 |
| **spotlight** | Full Fan | 21.4 | 28 | 125 | 500 | 60 | 25 | 750 | 200 | 30.6 | 29.7 | 44.7 |

# Dosimetric Measurements

## Air Kerma free-in-air

The air Kerma ($K_{air}$) free-in-air was measured using a PTW TN30013 Farmer-type ionization chamber (0.6 cc, PTW, Freiburg, Germany), calibrated for $K_{air}$. The ionization chamber was positioned at the isocenter of the CBCT imaging beam (Figure 1A). The air Kerma values for the default and modified protocols were calculated using:

$$K_{air} = M_{Qi} \cdot C_{T,P} \cdot N_{Kair,Qi} \qquad [1]$$

With:

- $M_{Qi}$: Ionization chamber charge reading for an entire CBCT scan.
- $C_{T,P}$ : Temperature and pressure correction factor, derived from the ambient conditions during measurements.
- $N_{K,air,Qi}$ : Calibration factor interpolated for the imaging beam quality *Qi*.

## Farmer Chamber calibration

To ensure accuracy, the calibration coefficients ($\boldsymbol{N_{K,air,Qi}}$) for the Farmer chamber were determined using beam qualities representative of the clinical CBCT protocols. First, the qualities of the clinical CBCT beams were determined through half-value layer (HVL) measurements performed in broad beam configuration with a Raysafe X2 R/F detector (Fluke Corporation, Glenwood, IL). Table 2 summarizes the clinical beam quality indices and HVLs (mm Al) for all clinical CBCT protocols. The calibration was then performed at the University of Madison Accredited Dosimetry Calibration Laboratory (UWADCL) using Tungsten anode beams (UW100-M, UW120-M and UW150-M) bracketing the clinical CBCT beams qualities. Table 3 summarizes the beam qualities, first half-value layers (HVL1), and calibration factors used for this study.

The beam qualities of the CBCT protocols were matched to the calibration beams based on the measured HVLs (Table 2) to interpolate the $N_{K,air,Qi}$ values. Each air Kerma measurement was repeated five times, and the standard uncertainty $u(K_{air})$ was calculated based on the variability across these readings.

*Table 2: beam quality measurements for the clinical combination of x-ray tube potential and filtering. Beam collimation was set to provide the maximum imaging field size.*

| Protocol | kVp | Filter | Collimation | | | | HVL 1 (mm Al) |
|---|---|---|---|---|---|---|---|
| | | | Y1 | Y2 | X1 | X2 | |
| Head | 100 | Ti + Full Fan | -10.7 | 10.7 | -10 | 14 | 7.5 |
| Pelvis/Thorax | 125 | Ti + Half Fan | -10.7 | 10.7 | -24.7 | 3.3 | 8.35 |
| Large Pelvis | 140 | Ti + Half Fan | -10.7 | 10.7 | -24.7 | 3.3 | 8.93 |
| Spotlight | 125 | Ti + Full Fan | -10.7 | 10.7 | -14 | 14 | 8.48 |

*Table 3: Farmer chamber calibration data*

| **Beam Quality** | **HVL1 (mm Al)** | $N_{K,air}$ **(Gy/C)** | **Filters (mm)** |
|---|---|---|---|
| UW150-M | 10.1 | 4.807E+07 | 2.83 Al + 0.28 Cu |
| UW120-M | 6.77 | 4.814E+07 | 3.0 Al + 0.1 Cu |
| UW100-M | 4.98 | 4.797E+07 | 4.77 Al |

## Cone Beam Dose Index (CBDI)

To complement air kerma ($K_{air}$) measurements, Cone Beam Dose Indices (CBDI) and weighted Cone Beam Dose Indices, ($CBDI_w$), [10] were determined using a 100 mm pencil Raysafe X2 CT detector (Fluke Corporation, Glenwood, IL) in PMMA cylindrical phantoms (Figure 1B). The CBDI values were measured at the center and periphery of the phantoms, and $CBDI_w$ was calculated as:

$$CBDI_w = \frac{1}{3} \cdot CBDI_{center} + \frac{2}{3} \cdot CBDI_{periphery} \qquad [2]$$

The relationships between $K_{air}$, CBDI, and $CBDI_w$ were evaluated for all protocols to establish conversion factors. These factors allow for the intercomparison of CBCT imaging protocols across different facilities.

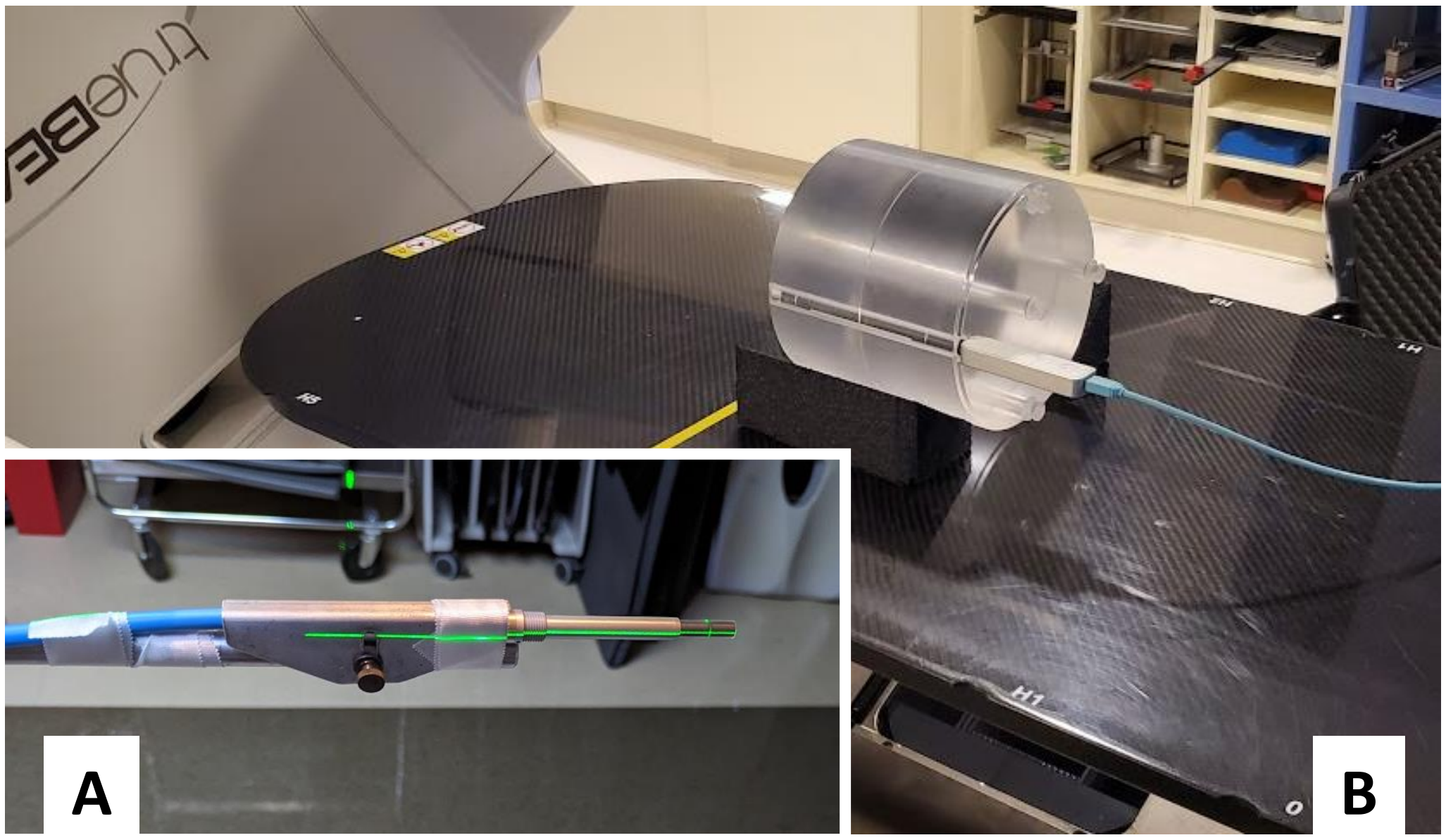

*Figure 1: (A) head protocol CBDI and isocenter air kerma ($K_{air}$) free-in-air measurement set-ups.*

## Image quality assessment

### Quantitative evaluation

Quantitative image quality metrics were assessed for all default and modified CBCT protocols using a Catphan604 phantom (PhantomLab, Salem, NY) [11] . Image analysis was performed with an automated in-house TG-142 QA tool based on Pylinac [12], ensuring consistent and efficient analysis across protocols. For low dose protocols where the automated toolkit was unable to extract quantitative metrics, scans were imported into Velocity (Varian Medical Systems, Palo Alto, CA) for manual analysis. When indicated, tolerances represent our institution values for monthly QA. The following metrics were evaluated:

1. HU Consistency and Uniformity:
    - HU Consistency: The CTP732 module, containing 9 plugs of different densities ranging from air to Teflon, was used to evaluate the HU values. The mean HU values of each plug were compared against baseline values established during annual QA. The tolerance was set at ±40 HU [13].
    - HU Uniformity: The CTP29 module was used to compare the mean HU value of four peripheral regions of interest (ROIs) (top, bottom, left, and right) to the central ROI. The tolerance for the difference between the central and peripheral ROIs was also set at ±40 HU [13]. The HU uniformity was characterized by the Uniformity Index (UI) and Integral Non Uniformity (IN) index, both calculated based on [14]:

    $$UI = 100 \cdot \frac{\overline{HU}_{periphery} - \overline{HU}_{center}}{\overline{HU}_{center} + 1000} \qquad [2]$$

    $$IN = \frac{\overline{HU}_{max} - \overline{HU}_{min}}{\overline{HU}_{max} + \overline{HU}_{min} + 2000} \qquad [3]$$

    Where $\overline{HU}$ represents the average pixel HU value in each ROI.

2. Low Contrast Visibility:
    - The low contrast visibility [15] was assessed using the CTP515 module. It combines low contrast resolution, contrast-to-noise ratio (CNR), and ROI size to estimate the detectability of the ROI. Visibility changes were correlated with exposure reductions.

3. High Contrast Resolution:

- The CTP528 module, containing 1–15 line pairs per centimeter, was used to measure the modulation transfer function (MTF). The MTF provides a relative measure of the system's ability to resolve high-contrast structures. The normalized MTF (nMTF) was computed as nMTF(f) = MTF(f)/$MTF_{max}$, where MTF(f) corresponds to the MTF at spatial frequency f, and $MTF_{max}$ corresponds to the maximum value of the MTF for a set of specific protocol, number of projections and mAs. The quantitative analysis focused on the MTF50 (spatial frequency at 50% of the maximum MTF) and MTF10 (spatial frequency at 10% of the maximum MTF), which are respectively representing the CBCT imaging system's effective resolution and limiting resolution.

4. Spatial Integrity:

- Spatial integrity was evaluated using the CTP528 module, which contains nodes spaced 50 mm apart. Slice thickness accuracy was measured using 23° ramps. Tolerances are +/- 1 mm from the nominal 50 mm node spacing and +/- 0.125 mm from a reference 2.0 mm slice thickness. Any value inside the tolerance is considered as "pass", while values outside the tolerance range is considered as 'fail'.

Example CBCT scans from the Catphan604 sensitometric (CPT732), high resolution (CPT528), and low contrast resolution (CPT515) modules are shown in Fig. 2.

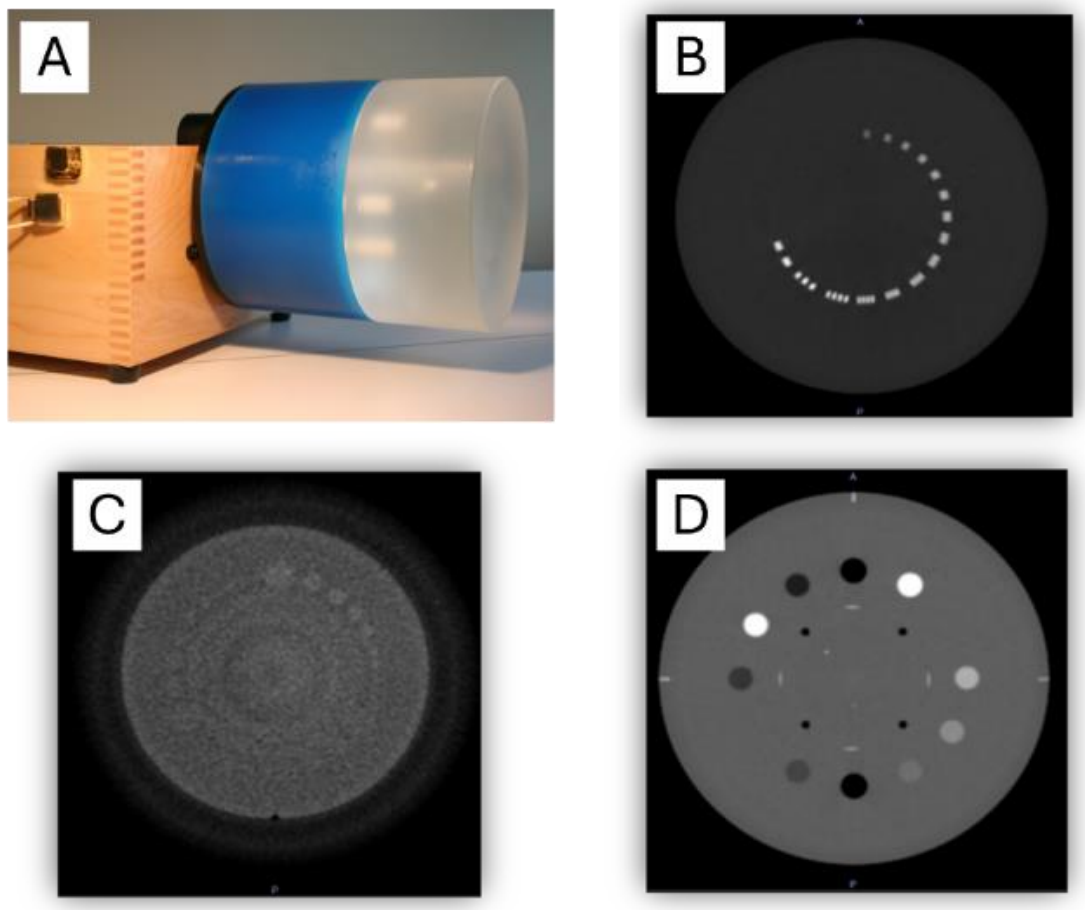

*Figure 2: (A) Catphan604, (B) high contrast, (C) low contrast and (D) sensitometry modules for the 50% frames/mAs THORAX protocol.*

## Statistical Analysis

The Pearson correlation coefficient (*R*) was calculated to assess the linear relationship between dose and five imaging parameters: HU uniformity, low contrast visibility, high contrast resolution, spatial integrity, and slice thickness. Statistical significance was determined using *p*-values, with $p<0.05$ considered statistically significant. The strength of correlations was classified as follows:

- $|R|\geq 0.7$: Strong correlation
- $0.5\leq|R|<0.7$: Moderate correlation
- $0.3\leq|R|<0.5$: Weak correlation
- $|R|<0.3$: No correlation

Correlation coefficients and their corresponding *p*-values were computed for each protocol to evaluate the impact of dose reductions on imaging quality. The statistical analysis was performed with IBM SPSS Statistics V.29 (IBM, Armonk, NY).

## Qualitative evaluation

Qualitative evaluation of image quality was performed for clinically relevant protocols with a Head (CIRS, Melbourne, FL), Thorax (RSD Radiology Support Devices Inc, Gardena, CA) and Pelvic (Brainlab AG, Munich, Germany) anthropomorphic phantoms (Figure 3). A multidisciplinary team, including a radiation oncologist, two physicists, and two radiation therapy technologists, reviewed the scans. They assessed the presence of aliasing artifacts, noise levels, and their potential impact on patient alignment for clinical use. Specific attention was paid to identifying thresholds at which dose reduction began to compromise image usability for patient alignment prior RT treatment.

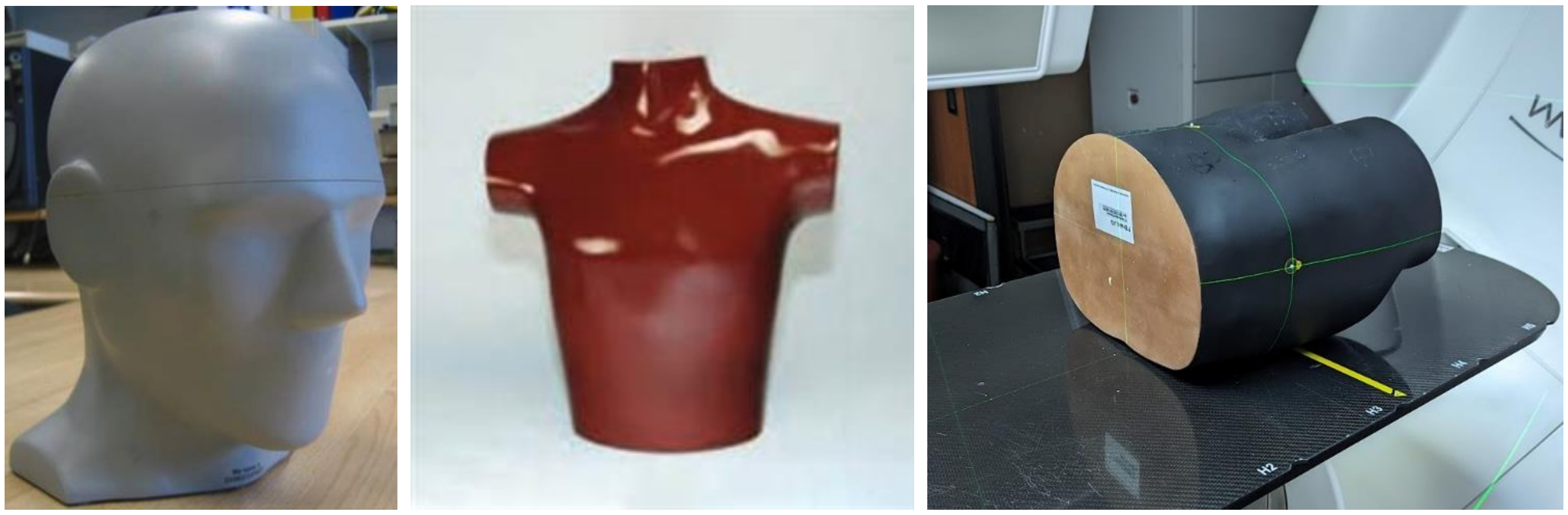

*Figure 3: (A) STEEV head phantom, (B) Thorax phantom, (C) pelvis phantom.*

### Clinical implementation

Optimized CBCT protocols were created from this study for clinical use. The selection of the low dose protocols was made keeping in mind the need for appropriate imaging quality for clinical use. Also, it was necessary to introduce these as small changes so that RTTs would not have to use a higher dose scan if the image quality was not appropriate for a specific patient. In addition, a pediatric protocol was created using this method for the treatment of a patient specific case with multiple previous RT treatments, which needed to minimize the CBCT protocol dose as low as possibly achievable while preserving the high contrast between bone and soft tissue for the treatment of bilateral mandibular metastases.

## Results

### Radiation Output Measurements

The measured air Kerma ($K_{air}$) values for the five default CBCT protocols are summarized in Table 1. A linear relationship was observed between $K_{air}$, Cone Beam Dose Index (CBDI), and weighted CBDI (CBDIw), enabling intercomparison of CBCT protocols. The normalized dose indices (CBDI/mAs, CBDIw/mAs, and $K_{air}$/mAs) and conversion coefficients between CBDI, $CBDI_w$, and $K_{air}$ were calculated for each protocol and are listed in Table 4. The time required for $K_{air}$ measurements was more than ten times shorter than for CBDI measurements, underscoring the practicality of this method for routine use in radiotherapy centers.

*Table 4: Normalized CBDI, $CBDI_w$, $K_{air}$ and conversion coefficients between CBDI, $CBDI_w$ and $K_{air}$.*

| Protocol | CBDI (mGy/mAs) | $CBDI_w$ (mGy/mAs) | K_air (mGy/mAs) | CBDI / $K_{air}$ | $CBDI_w$ / $K_{air}$ |
|---|---|---|---|---|---|
| **Head** | 0.022 | 0.021 | 0.034 | 0.633 | 0.614 |
| **Thorax** | 0.013 | 0.018 | 0.063 | 0.208 | 0.280 |
| **Pelvis** | 0.012 | 0.017 | 0.059 | 0.211 | 0.283 |
| **Pelvis large** | 0.017 | 0.022 | 0.077 | 0.221 | 0.290 |
| **Spotlight** | 0.040 | 0.039 | 0.059 | 0.687 | 0.665 |

### Quantitative Image Quality Assessment

### HU constancy and Uniformity

HU constancy and HU Uniformity remained stable across most modified protocols. HU differences between central and peripheral ROIs were within the ±40 HU tolerance for all modified protocols. The HU Constancy results showed differences from baseline values within the tolerance of ±40 HU. Exceptions were observed for the constancy of LARGE PELVIS modified protocols with full number of projections and with single-frame exposures below 50%, where constancy deviated significantly for high density materials. Figure 4 presents HU constancy results for the HEAD, PELVIS, LARGE PELVIS, SPOTLIGHT and THORAX protocols, expressed as the average difference between the default and modified protocols for varying mAs. The error bars represent +/-1 standard deviation. The lower the dose, the higher the variance in HU difference. Each graph separates dose reduction obtained by: (1) reducing the number of projections used for reconstruction, keeping the single frame exposure as provided by the default protocols; or (2), by reducing the single frame exposure relative to the default protocols and keep the maximum number of frames for reconstruction. Material specific differences tend to be larger for high density material (Teflon, Bone_50%, Bone_20% and Delrin), but only weak and non-significant correlations were observed between dose reduction and HU differences across protocols. Stratifying by dose reduction method revealed underlying strong correlations for the Pelvis, Large Pelvis, Spotlight protocols.

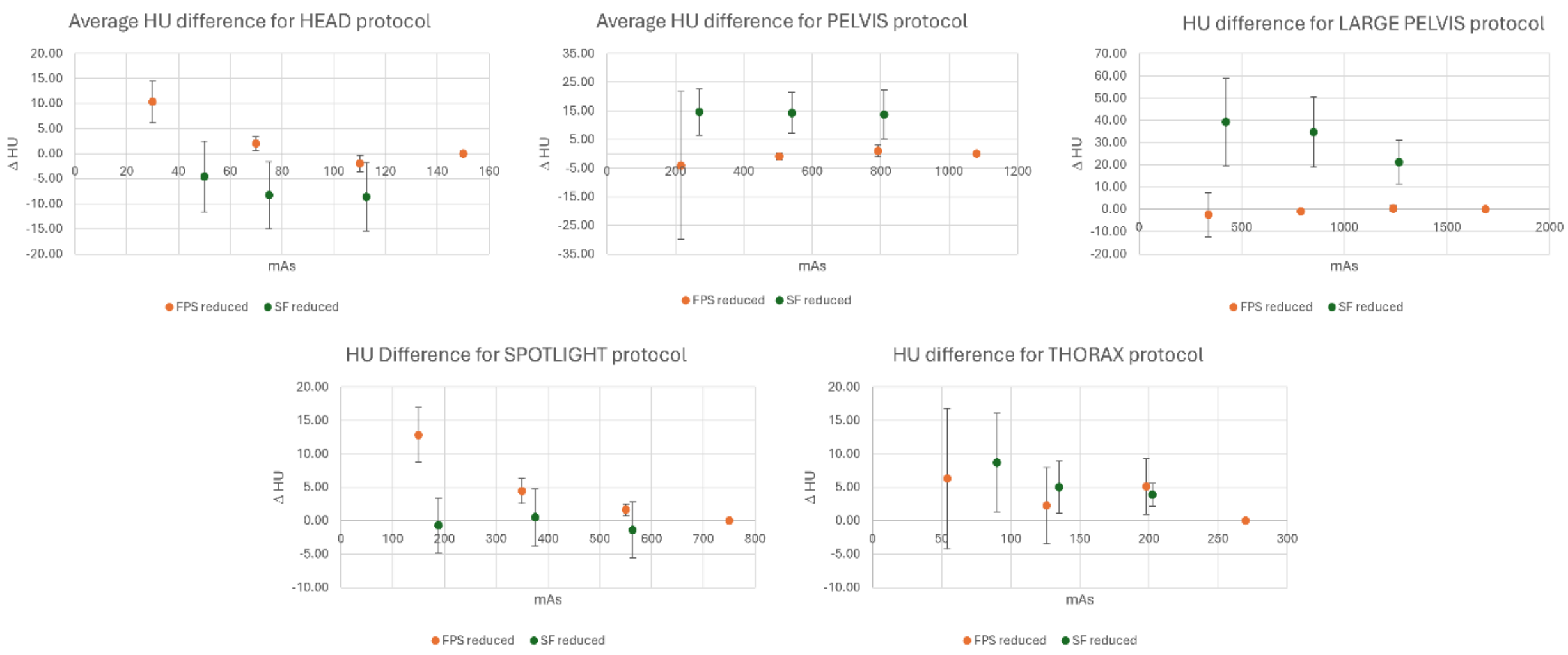

*Figure 4: Variation of average HU difference for each protocol with mAs.*

## Low contrast visibility

Overall, image visibility decreased progressively with lower exposures levels. Figure 6 illustrates the trends in low contrast visibility across all protocols. The HEAD (blue) and THORAX (orange) protocols exhibited the lowest visibility, reflecting their low exposure range optimized for small volumes (head) and low density tissue (lung in THORAX).

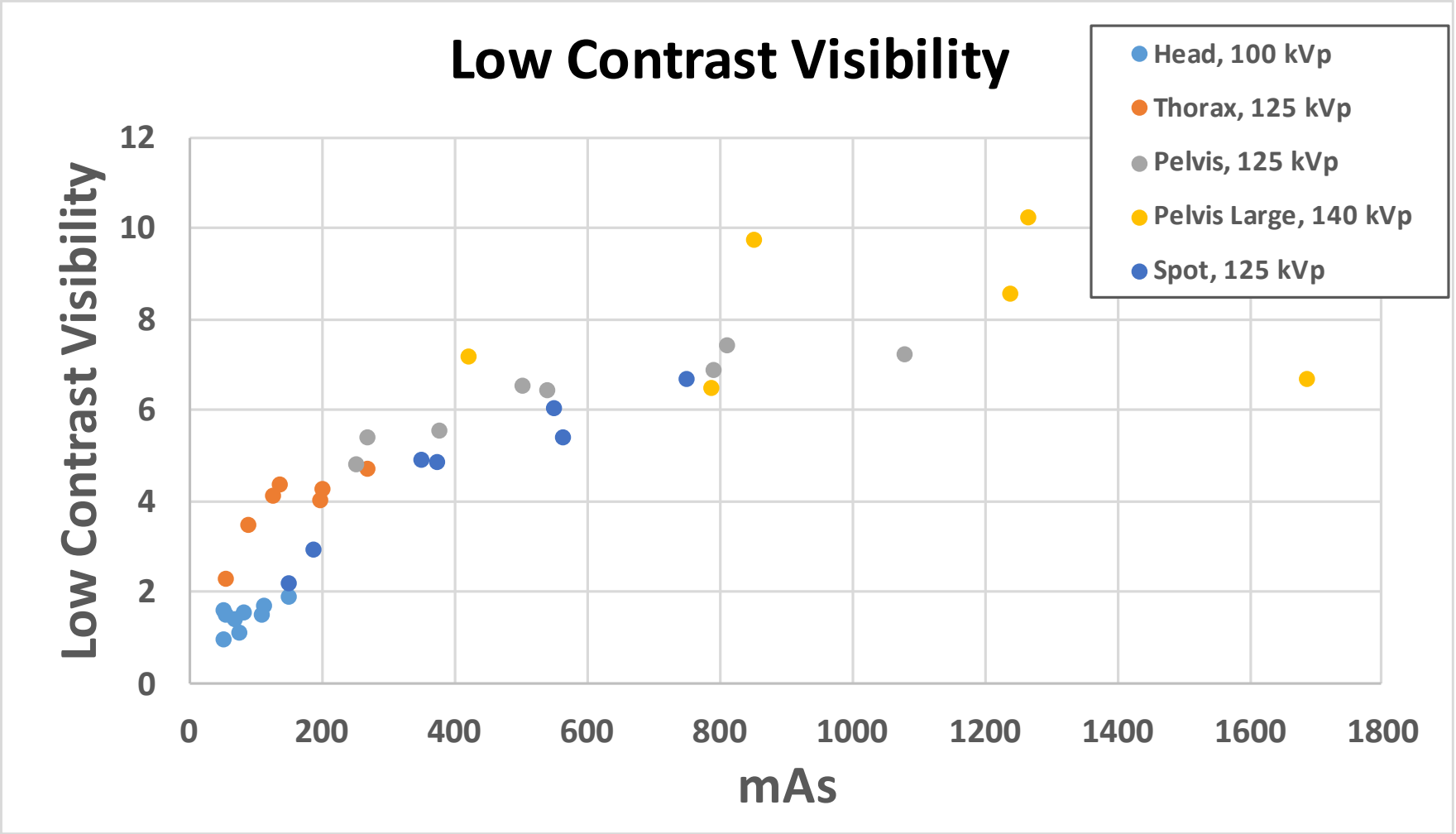

*Figure 5: Low contrast visibility for all protocols. The low contrast visibility decreases with exposure and beam quality. The Head and Spotlight protocols will be more affected by a reduction in x-ray source output.*

## High contrast resolution

High-contrast resolution, measured via relative Modulation Transfer Function (rMTF), remained stable across all exposure levels, as shown in Figure 6. This finding indicates that dose reduction does not compromise high-contrast detail, ensuring that critical structures with strong density differences remain clearly visible.

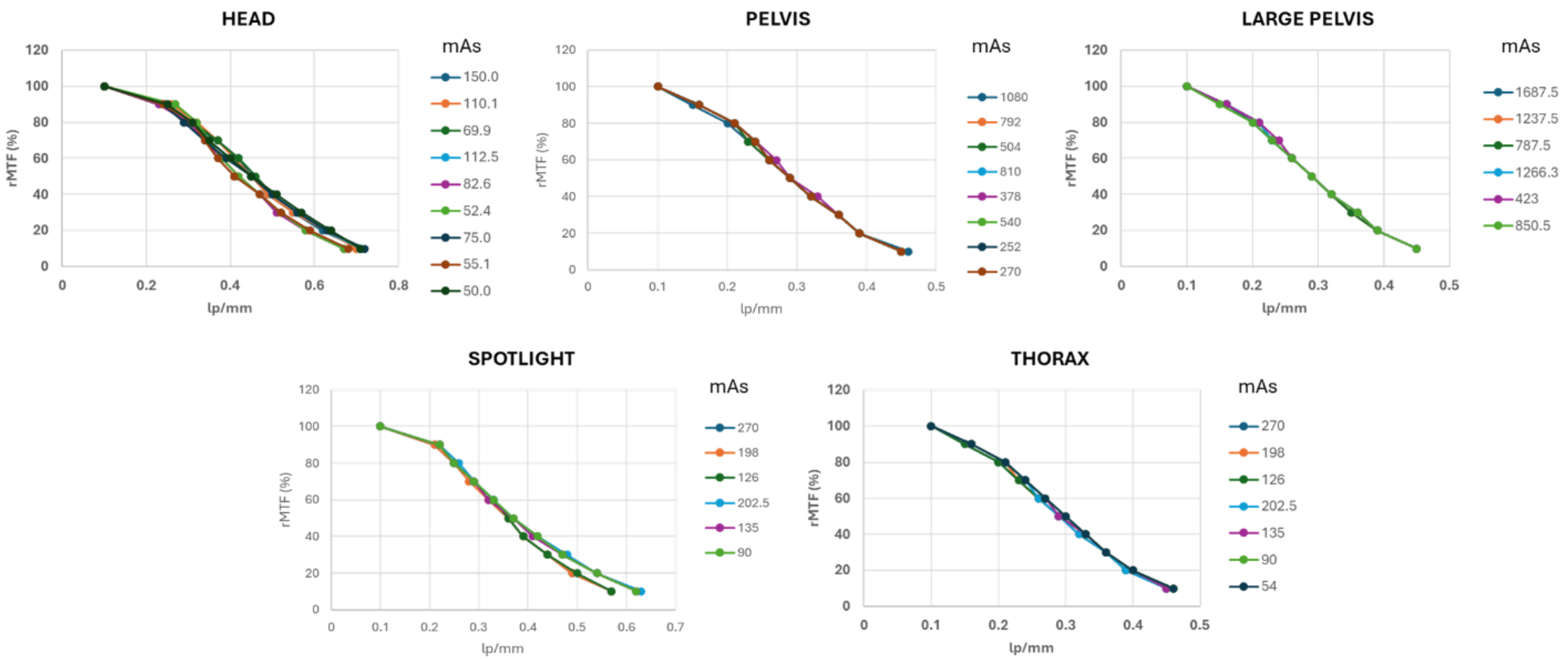

*Figure 6: Relative Modulation Transfer Function (rMTF) for each CBCT protocols, illustrating the absence of correlation between high contrast resolution and reduction in exposure.*

## Spatial integrity (in-plan and slice thickness)

Spatial integrity parameters were unaffected by exposure reductions, with in-plane distances and slice thickness measurements remaining well within their respective tolerance ranges for all modified protocols.

## Correlation Analysis

Table 5 summarizes the Pearson correlation coefficients (*R*) and statistical significance (*p*-values) for each protocol and parameter. Significant correlations (*p*<0.05) highlight the relationships between dose and image quality parameters, with varying degrees of sensitivity observed across protocols:

1. **HU Uniformity**: Moderate negative correlations were observed for HEAD (-0.685, p < 0.05) and PELVIS (-0.665, not significant), indicating a tendency for HU uniformity to degrade with dose reductions. However, this trend was not statistically significant across all protocols.
2. **Low Contrast Visibility**: Strong positive correlations (R > 0.7, p < 0.05) were observed for most protocols (HEAD, THORAX, PELVIS, and SPOTLIGHT), reinforcing its sensitivity to dose variations. However, Large Pelvis (R = 0.093, not significant) showed no clear correlation, suggesting that visibility may be less affected by dose reductions in this protocol.
3. **High Contrast Resolution (rMTF50 and rMTF10)**: MTF50 showed moderate correlations for some protocols (HEAD: 0.663, p < 0.05), while MTF10 exhibited weaker or no significant correlations across all protocols. These findings indicate that high-contrast resolution remains stable despite dose reductions.
4. **Spatial Integrity**: Correlations varied across protocols, with some moderate associations (e.g., PELVIS: -0.667, THORAX: -0.673) suggesting minor dose-related changes. However, no protocol demonstrated a consistent or significant dose-dependent trend. The small relative variations observed between protocols reinforce that spatial integrity is preserved across dose levels.
5. **Slice Thickness**: Similar to the spatial integrity, minor variations in slice thickness were observed, but no strong or consistently significant correlations were identified. This suggests that dose reduction had a minimal impact on this parameter.

*Table 5: Correlation analysis of image quality metrics across CBCT protocols. The Pearson correlation coefficient (R) is reported for each parameter within each protocol. A * next to the R value indicates statistical significance (p < 0.05). The color scale in the table represents the strength of the correlation: Green (Strong correlation with |R| ≥ 0.7), Orange (Moderate correlation (0.5 ≤ |R| < 0.7), Brown (Weak 0.3 ≤ |R| < 0.5).*

| *PROTOCOL* | Visibility | Uniformity | Integral Non Uniformity | Line average | Slice thickness | MTF50 | MTF10 |
|---|---|---|---|---|---|---|---|
| *HEAD* | 0.754 * | -0.685 * | -0.378 | 0.024 | 0.663 * | 0.515 | 0.498 |

| Correlation |
|---|
| Strong |

| | | | | | | | | |
|---|---|---|---|---|---|---|---|---|
| ***PELVIS*** | 0.906 | * | -0.665 | 0.433 | -0.667 | -0.626 | 0.404 | 0.417 |
| ***LG PELVIS*** | 0.093 | | -0.162 | 0.536 | -0.180 | 0.196 | 0.548 | 0.548 |
| ***THORAX*** | 0.816 | * | -0.593 | 0.062 | -0.673 | 0.023 | -0.752 | -0.500 |
| ***SPOTLIGHT*** | 0.952 | * | -0.008 | -0.165 | -0.075 | 0.078 | 0.541 | -0.093 |

Moderate
Weak
None
* ($p < 0.05$)

## Qualitative evaluation

Figure 7 presents CBCT scans of anthropomorphic phantoms acquired using the HEAD, THORAX, and PELVIS protocols, evaluated by the evaluation team. The first column (A1–F1) shows images from the default protocols. Rows A, C, and E illustrate the impact of reducing the number of projections used for FDK reconstruction, while rows B, D, and F demonstrate the effect of decreasing the single-frame mAs while maintaining the maximum number of projections. The scan parameters, number of projections, and corresponding CBDI and air kerma ( $K_{air}$) dose metrics are indicated above each image. The review of CBCT scans by the evaluation team revealed the following:

1. Aliasing artifacts became more pronounced with fewer projections (Fig. 7, image A4, C4 and E4), particularly for HEAD and SPOTLIGHT protocols with their smaller angular ranges (200° arcs).
2. **Noise levels** increased as single-frame exposure decreased but remained clinically acceptable in most cases, except at the lowest exposure settings.
3. Dose reductions of 50–80% maintained clinically acceptable image quality for patient alignment across most protocols.

## Optimized Clinical Protocols

The optimized protocols selected for clinical implementation are summarized in Table 6. These protocols achieved cumulative exposure reductions of 50–80% compared to the default protocols, while retaining adequate image quality for clinical use. Example axial CBCT slices of anthropomorphic phantoms for each optimized protocol demonstrate minimal artifacts and noise interference. Figure 7  illustrates the clinical implementation of reduced dose HEAD CBCT protocols (22% dose reduction) for IGRT of Head and Neck case and of a reduced dose PELVIS Protocol (48% dose reduction) for the treatment of a rectal cancer patient.

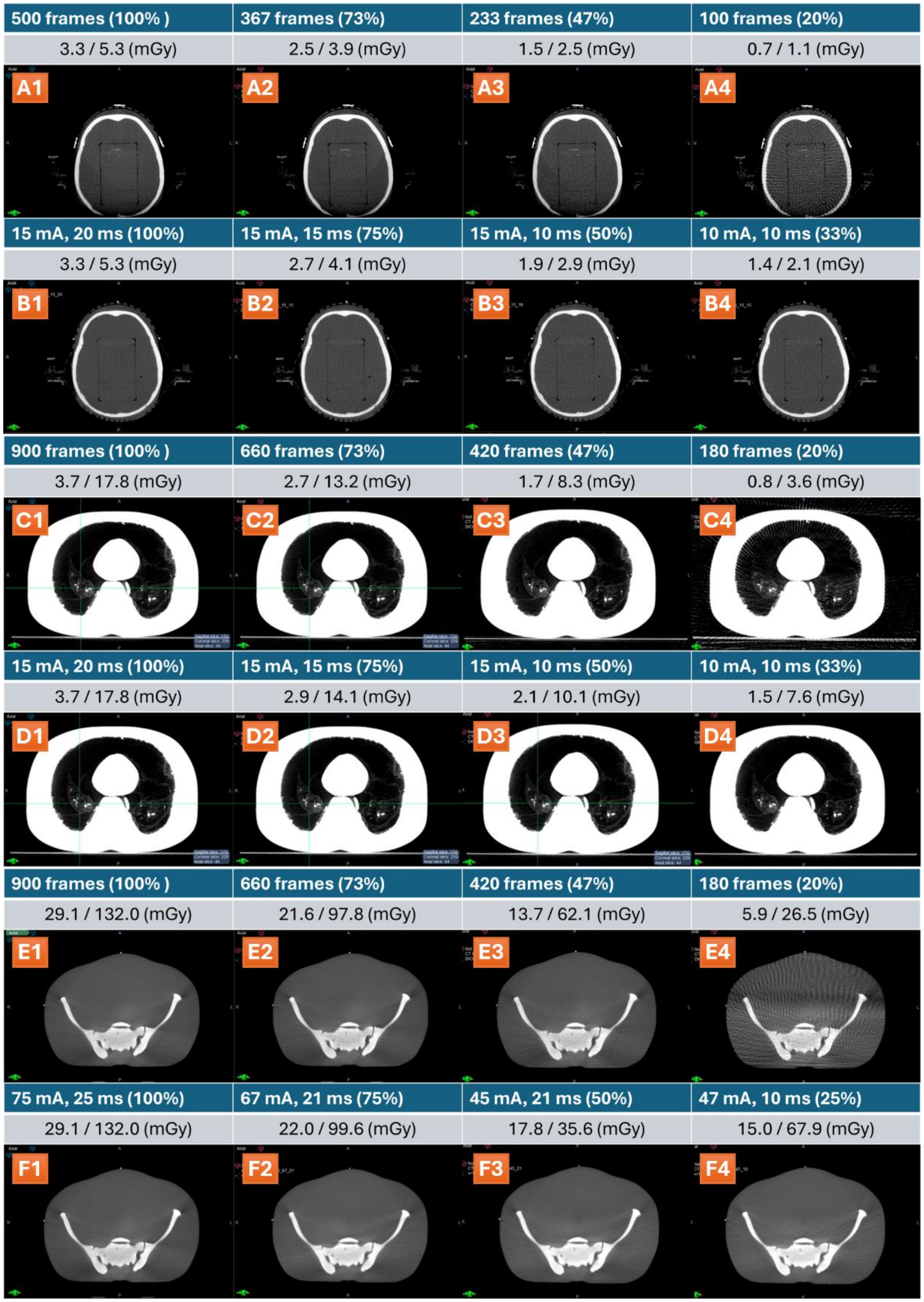

*Figure 7: CBCT images of anthropomorphic phantoms acquired using HEAD (A, B), THORAX (C, D), and PELVIS (E, F) protocols to evaluate image quality under different dose reduction strategies. The first column (A1–F1) displays images from the default protocols. Rows A, C, and E illustrate the impact of dose reduction by decreasing the number of projections used for FDK reconstruction, while rows B, D, and F demonstrate the effect of reducing the single-frame mAs while maintaining the maximum number of projections. The scan parameters, number of projections, and corresponding CBDI/Kair values are indicated above each image.*

*Table 6: Optimized CBCT protocols for **Head, Thorax, Pelvis, Large Pelvis, and Spotlight** selected for clinical use. The table presents the cumulative exposure (mAs), number of projections used for reconstruction, and the corresponding $K_{air}$ (mGy), along with the relative dose reduction compared to the original default protocols.*

| Protocol | Exposure (mAs) | Projections | $K_{air}$ (mGy) | Relative Dose reduction |
|---|---|---|---|---|
| **HEAD** | 112.5 | 500 | 4.13 | 78% |
| **THORAX** | 135 | 900 | 11.11 | 57% |
| **PELVIS** | 540 | 900 | 33.11 | 52% |
| **LARGE PELVIS** | 850.5 | 900 | 67.90 | 51% |
| **SPOTLIGHT** | 563.5 | 500 | 33.87 | 76% |

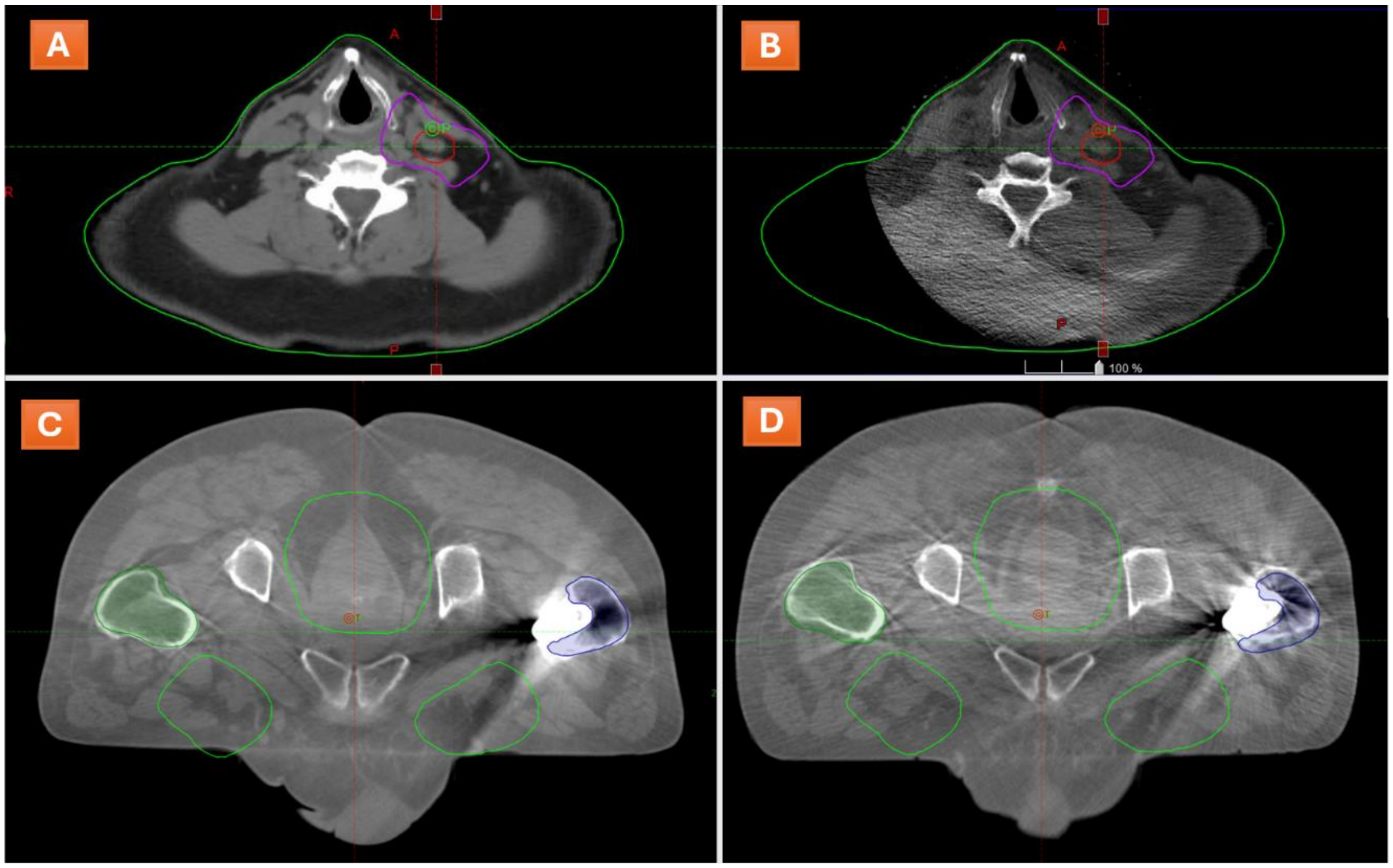

*Figure 8: clinical example of low dose protocols for a head and neck patient and a rectal cancer patients. (A) Head and Neck planning CT, (B) CBCT acquired during the first fraction. (C) patient imaged with default PELVIS CBCT vs (D) Low dose PELVIS CBCT.*

## Pediatric CBCT Dose Optimization for Neuroblastoma IGRT

For a pediatric patient case, the optimized CBCT protocols achieved substantial dose reductions while maintaining accurate patient alignment. The patient was an 11-year-old boy with metastatic neuroblastoma in the bilateral mastoid region, scheduled to receive 21.6 Gy in 12 fractions of 1.8 Gy, with daily CBCT for IGRT. Given the patient's history of multiple radiation treatments, the CNS radiation oncologist requested a very low-dose CBCT protocol for IGRT.

Since the CTDI measurement equipment was unavailable, dose estimation was performed using the air kerma calibrated Farmer chamber. A pediatric-specific HEAD protocol was developed with the following modifications: (1) the tube potential was reduced from 100 kVp to 80 kVp (compared to the default HEAD protocol); (2) a total of eight dose-reduced protocols were evaluated, incorporating reductions in both the number of projections and single-frame mAs. Phantom studies were conducted to assess image quality at reduced imaging doses. Dose was measured as air kerma (Kair) relative to the default HEAD protocol. The dose measurements and phantom studies were completed in less than two hours. Two optimized protocols were selected for review by the CNS radiation oncologist: (1) HEAD_PEDS with 500 projections and 75% dose reduction; and (2) HEAD_PEDS_LD, with 367 projections and 88% dose reduction. The CNS oncologist selected HEAD_PEDS_LD, justifying the choice based on the need for patient alignment using only bony anatomy.

Figure 8 presents phantom images of the default HEAD protocol (Fig. 8A) alongside HEAD_PEDS_LD (fig. 8B), as well as the patient's CBCT scans from the first treatment fraction, demonstrating the left and right mastoid targets (Fig. 8C and 8D). This case underscores the potential of optimized CBCT protocols to minimize radiation exposure in pediatric and other sensitive patient populations.

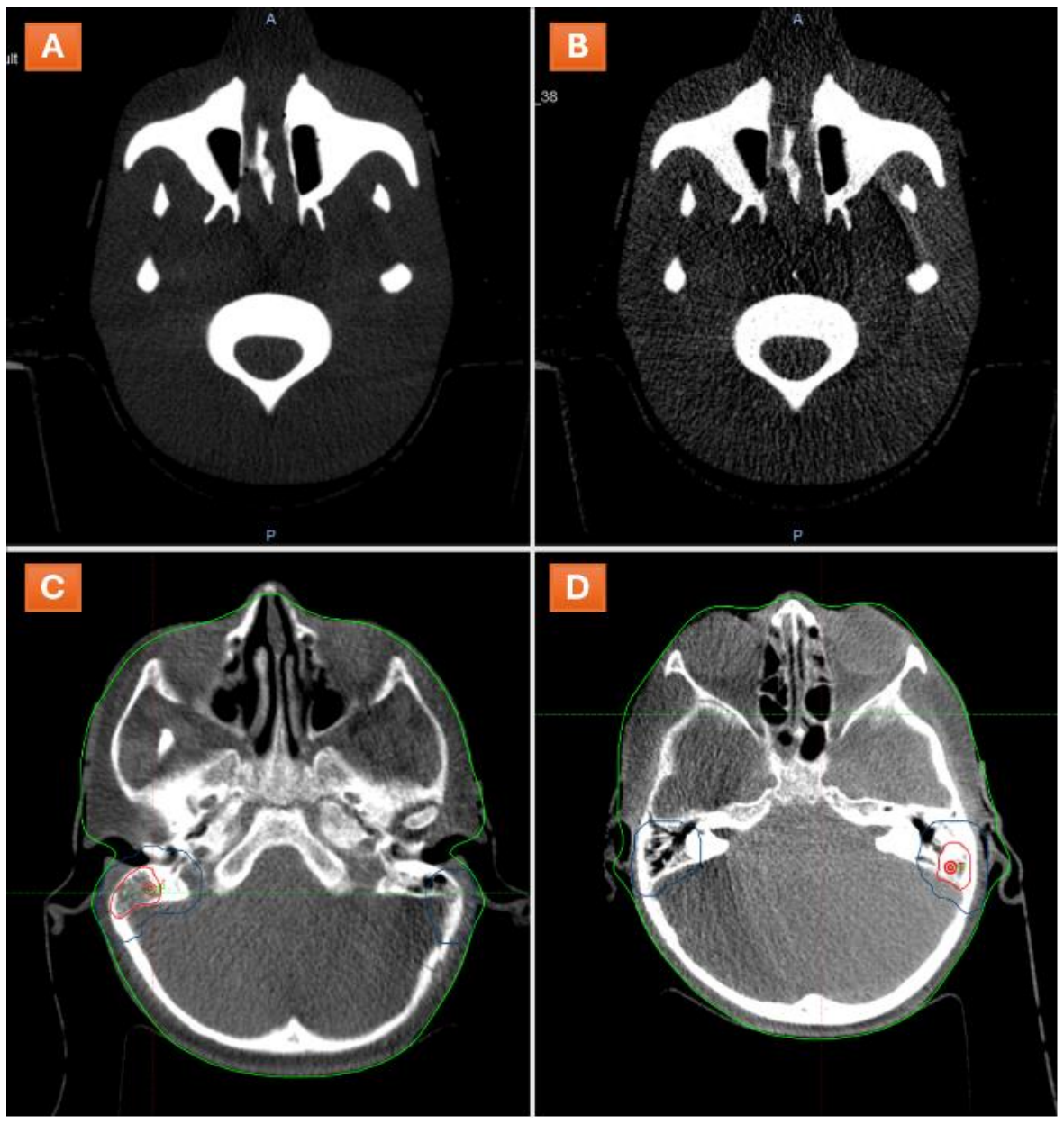

*Figure 9: Comparison of CBCT images for a pediatric patient undergoing IGRT. Phantom study: (A) Default HEAD protocol; (B) Optimized **HEAD_PEDS_LD** protocol with an 88% dose reduction, maintaining sufficient image quality for alignment. Patient images: (C, D) first fraction CBCT images demonstrating alignment based on bony anatomy, with the left (C) and right (D) mastoid target regions outlined. This case highlights the feasibility of significant dose reduction while preserving clinical utility in pediatric patients.*

## Discussion

The optimization of cone-beam computed tomography (CBCT) protocols in image-guided radiotherapy (IGRT) is critical for reducing cumulative imaging doses while maintaining clinically acceptable image quality. This work demonstrates a practical approach to CBCT dose optimization that can be implemented in radiotherapy clinics using readily available equipment. The study establishes that reducing both the number of projections and single-frame exposures can significantly lower CBCT dose without compromising essential imaging quality metrics. The linear relationship between $K_{air}$, CBDI, and $CBDI_w$ provides a reliable and efficient method for evaluating dose reductions across different protocols. This approach is practical for clinics without access to advanced diagnostic physics tools, as it reduces measurement time by a factor of 10 compared to CBDI measurements.

In terms of Image Quality Metrics, HU linearity and uniformity remained within clinical tolerances for most protocols, indicating that dose reductions can be achieved without significant degradation in quantitative imaging quality. Low contrast visibility, which showed the strongest correlation with exposure reduction, emerged as the primary limiting factor in dose optimization. While visibility decreased at lower exposures, clinically acceptable levels were maintained for reductions up to ~90% in certain protocols. Although soft tissue visualization is significantly impacted by dose reduction, IGRT relying solely on bony anatomy remains unaffected, as demonstrated by the successful implementation of the pediatric-specific protocol.

Qualitative assessments by a multi disciplinary team confirm that dose reductions of 50–80% retain sufficient image quality for patient alignment in clinical practice. However, care must be taken to minimize artifacts when reducing projections, particularly for protocols with smaller angular ranges, such as HEAD and SPOTLIGHT. The input of Radiation Therapy Technologists (RTTs) as end users of IGRT is essential in evaluating the clinical usability of low-dose protocols. Additionally, phantom studies provide a systematic approach to testing low-dose protocols before clinical implementation, preventing the need to use patients as "test subjects." for optimizing IGRT protocols [16]. Our proposed approach reduces the risk of acquiring an initial low-dose image with limited clinical usefulness, which could necessitate a higher-dose CBCT scan, ultimately leading to increased imaging dose to the patient.

Previous studies have examined CBCT dose optimization in IGRT, but most focused on either dose reduction or image quality assessment separately. Yan et al. [17] provided a comprehensive study of both in the context of CBCT reconstruction algorithm development, but their investigation was limited to a full Fan acquisition mode similar to our HEAD and SPOTLIGHT protocols. Our findings extend their work by demonstrating that dose reductions of up to 80% remain clinically viable across multiple acquisition modes, including half-fan protocols used for THORAX and PELVIS imaging.

Additionally, prior research has explored the trade-off between dose and image quality in manufacturer-provided protocols [18-20] but few studies have systematically tested site-specific protocol modifications tailored to radiotherapy applications. Our study bridges this gap by developing an institution-specific CBCT dose optimization method, demonstrating its feasibility across multiple treatment sites, including a successful pediatric implementation for neuroblastoma IGRT.

Furthermore, while AAPM TG-180 [2] and other reports [8, 9], outline dose management strategies, they require diagnostic dosimetry equipment, which limits their applicability in standard radiotherapy clinics. Our proposed method, relying on $K_{air}$ measurements instead of traditional $CTDI_w$ assessments, offers a practical alternative that reduces measurement time by a factor of 10 while maintaining robust dose accuracy.

## Limitations and future directions

While this study provides a practical framework for CBCT dose optimization, several limitations should be considered. The optimized parameters were developed specifically for Varian TrueBeam systems and may require adaptation for other linac models or manufacturers. Additionally, the qualitative nature of image assessments, despite being conducted by a multidisciplinary team, introduces some degree of subjective variability. At very low exposure levels, increased noise and artifacts may limit clinical usability, particularly in protocols requiring high soft-tissue contrast.

Future work should focus on expanding this methodology to other IGRT platforms offering CBCT imaging. Establishing a multi-institutional collaboration would facilitate broader validation and enable cross-comparison of dose levels across different IGRT modalities. Additionally, further research should explore advanced image reconstruction algorithms, such as iterative CBCT (iCBCT) reconstruction on Varian TrueBeam systems [21], to enhance image quality at lower dose levels.

Efforts to standardize CBCT dose reference levels (DRLs) in IGRT are already underway through international initiatives, including the ICRP TG-116 mentee network and recent UK-based CBCT surveys [22]. These studies provide a framework for establishing national DRLs, a critical step in ensuring consistent imaging dose management across radiotherapy centers. Aligning future dose optimization methods with these ongoing initiatives will help develop comprehensive guidelines that are clinically feasible and globally applicable.

## Conclusion

This study presents a practical and clinically feasible CBCT dose optimization method that can be implemented by radiotherapy physicists using readily available measurement equipment. The approach allows for substantial dose reductions while preserving clinically acceptable image quality, making it suitable for routine clinical use. By establishing a linear relationship between $K_{air}$ and CBDI, this method enables inter-institutional comparison of CBCT protocols, facilitating standardization across different centers. The findings demonstrate that reducing both the number of projections and single-frame exposure can lower CBCT dose without compromising spatial integrity, HU uniformity, or high-contrast resolution. However, low-contrast visibility remains the primary limiting factor, requiring careful consideration when optimizing protocols. Importantly, the successful implementation of a pediatric-specific low-dose protocol highlights the potential of this approach in minimizing radiation exposure for sensitive populations. Future efforts should focus on expanding this methodology to other IGRT platforms, incorporating advanced image reconstruction techniques, and establishing standardized dose reference levels (DRLs) for CBCT in IGRT. These steps will further enhance radiological protection practices while maintaining the diagnostic accuracy essential for precise treatment delivery in radiotherapy.

## Acknowledgments

The authors would like to thank Dr. Colin Martin and all members of ICRP TG-116 “Radiological Protection Aspects of Imaging in Radiotherapy” for the on-going fruitful discussions on the optimization of radiological protection in IGRT.